\documentstyle[prl,aps]{revtex}
\input epsf

\begin{document}


\title{Space of kink solutions in $SU(N)\times Z_2$}

\author{
Levon Pogosian}
\address{
Theoretical Physics, The Blackett Laboratory, Imperial College,\\
Prince Consort Road, London SW7 2BZ, U.K.. }
\author{
Tanmay Vachaspati}
\address{
Department of Astronomy and Astrophysics,
Tata Institute of Fundamental Research,\\
Homi Bhabha Road,
Colaba, Mumbai 400005, India\\
and\\
Department of Physics,
Case Western Reserve University,\\
10900 Euclid Avenue, Cleveland, OH 44106-7079, USA.}

\wideabs{
\maketitle

\begin{abstract}
\widetext We find $(N+1)/2$ distinct classes (``generations'') of
kink solutions in an $SU(N)\times Z_2$ field theory. The classes
are labeled by an integer $q$. The members of one class of kinks
will be globally stable while those of the other classes may be
locally stable or unstable. The kink solutions in the $q^{th}$
class have a continuous degeneracy given by the manifold
$\Sigma_q=H/K_q$, where $H$ is the unbroken symmetry group and
$K_q \subseteq H$ is the group under which the kink solution 
remains invariant.
The space $\Sigma_q$ is found to contain incontractable two spheres
for some values of $q$, indicating the possible existence of certain
incontractable
spherical structures in three dimensions. We explicitly construct
the three classes of kinks in an $SU(5)$ model with quartic potential
and discuss the extension of these ideas to magnetic monopole solutions
in the model.
\end{abstract}
\pacs{} }

\narrowtext

\section{Introduction}
\label{introduction}

It is relatively easy to determine if a field theory with
spontaneous symmetry breaking admits topological defects. If the
asymptotic field configuration is topologically non-trivial, the
interior field configuration must have a topological defect.
However, there can be a large class of asymptotic field
configurations, all having the same topological characteristics.
Which of the many different boundary conditions with given
topology should one use when trying to find a topological defect
solution?

We will restrict our attention to the simplest kind of topological
defects, namely kinks in one spatial dimension. However the field
theories we will consider are rather general, having symmetry
groups $SU(N)\times Z_2$ with $N$ being an odd integer. The field
content will be a scalar field $\Phi$ transforming in the adjoint
representation of $SU(N)$, and the $Z_2$ takes $\Phi$ to $- \Phi$.
The potential of the field theory is taken to be such that it
gives a vacuum expectation value of $\Phi$ that breaks the
symmetry spontaneously to $H = [SU((N+1)/2)\times
SU((N-1)/2)\times U(1)]/C$, where $C=Z_{(N+1)/2}\times Z_{(N-1)/2}$ is the
center of $SU((N+1)/2)\times SU((N-1)/2)$;
other than having this property the
potential is not restricted in any way. The vacuum manifold of the
theory is disconnected because the $Z_2$ is broken down completely
by the vacuum expectation value. Hence there are topological kinks
in the theory.

Suppose we want to find the explicit solution for these kinks. 
Let $\Phi (x=-\infty ) = \Phi_-$ and $\Phi (x=+\infty ) =
\Phi_+$. Then, to obtain a topological defect,
the only constraint is that $\Phi_+$ and $\Phi_-$
should lie in distinct topological sectors of the vacuum manifold.
In fact, if $\Phi_+$ is a choice, $U\Phi_+ U^\dagger$ for $U\in
SU(N)$ is also a valid choice. In \cite{PogVac00} it
was shown that the $SU(5)\times Z_2$ kink with $\Phi_+=-\Phi_-$ is
unstable to small perturbations and that there exists a
stable domain wall solution of lower energy corresponding to a
different choice of $\Phi_+$. These results were generalized to
$SU(N)\times Z_2$ in \cite{Vac01} where the concept of different
classes of kink solutions was introduced. Given a kink solution,
the rest of the solutions from the same class can be constructed by
applying global gauge transformations from the coset space $H/I$
where $H$ is the unbroked symmetry group and $I \subseteq H$ is the 
``internal'' symmetry
group that leaves the original kink solution invariant. One such
class of solutions was constructed in \cite{Vac01}, however,
several questions of relevance were left unanswered. Will
there exist a kink solution for any choice of $\Phi_+$? Are the
different solutions really distinct? How many distinct solutions
can one obtain? Are these solutions stable? We will answer these
questions in this paper.

In Sec. \ref{kinksolutions} we will show that not all choices of
$\Phi_+$ lead to kink solutions and we find that we must have
$[\Phi_+, \Phi_-]=0$ in order for a solution to exist. This leads
to a finite, discrete set of topological boundary conditions that
can yield distinct kink solutions. Each boundary condition
determines a class of continuously degenerate kink solutions in
the model. Surprisingly, we also find that there are
non-topological kink solutions for which the boundary conditions
do not lie in distinct topological sectors. These solutions can
also be classified and counted. We then find the manifold that
describes the continuous degeneracy of every class. This manifold
has non-trivial topological properties which suggests that certain
closed domain walls are incontractable. In Sec. \ref{su5kinks} we
consider
the specific example of an $SU(5)$ model with a quartic potential
and construct the topological and non-topological kink solutions
explicitly. In this case we also analyze the stability of the kink
solutions in the three different classes. There is one globally
stable class of solutions; another is locally stable for some
parameters; the remaining classes are unstable for our
choice of potential.

In Sec. \ref{su5mm} we discuss the extension of our results on
domain walls to $SU(5)$ magnetic monopoles. With fixed asymptotic
field configurations, our findings suggest that there should exist
three generations of fundamental $SU(5)$ magnetic monopole
solutions. We summarize our results in Sec. \ref{conclusions}.

\section{Kink boundary conditions}
\label{kinkbc}

The Lagrangian of our (1+1 dimensional) model is:
\begin{equation}
L = {\rm Tr} (\partial_\mu \Phi )^2 - V(\Phi ) \ .
\label{lagrangian}
\end{equation}
$V(\Phi )$ is a potential invariant under
\begin{equation}
G\equiv SU(N) \times Z_2 \ , \label{originalsymm}
\end{equation}
$N$ is taken to be odd, and the parameters in $V$ are such that
$\Phi$ has an expectation value that can chosen to be
\begin{equation}
\Phi_0 = \eta \sqrt{2 \over {N(N^2-1)}}
                  \pmatrix{n{\bf 1}_{n+1}&{\bf 0}\cr
                      {\bf 0}&-(n+1){\bf 1}_n\cr} \ ,
\label{phi0}
\end{equation}
where ${\bf 1}_p$ is the $p\times p$ identity matrix and $\eta$ is
an energy scale determined by the minima of the potential $V$.
Such an expectation value spontaneously breaks the symmetry down
to:
\begin{equation}
H = [SU(n+1)\times SU(n)\times U(1)]/C \label{unbrokensymm} \ ,
\end{equation}
where we have defined
\begin{equation}
N\equiv 2n+1 \ , \label{littlen}
\end{equation}
with $n \ge 1$ being an integer. The exact form of $V(\Phi )$ will
not be important for most of our analysis. However, it does play a
role in the stability of solutions and then we will choose it to
be a quartic polynomial in $\Phi$.

If $\Phi (x=-\infty )=\Phi_-$, then $\Phi (x=+\infty) = \Phi_+ = -
U\Phi_- U^\dagger$ for any $U \in SU(N)$ implies that the boundary
conditions are topologically non-trivial. For example, if $U\in
H$, the symmetry group that leaves $\Phi_-$ invariant, then
$\Phi_+ = -\Phi_-$. The
first question we ask is: for a fixed $\Phi_-$, for what choices
of $\Phi_+$ can we obtain kink solutions? As we shall now see, for
a solution to exist, we must necessarily choose $\Phi_+$ such that
$[\Phi_+ , \Phi_- ]=0$.

In Appendix \ref{appendixa} we will prove the stronger result that
if $\Phi_k (x)$ is a solution then $[\Phi_\pm , \Phi_k (x)] =0$.
Here we will give a qualitative argument in support of this
statement. Once the boundary condition at $x=-\infty$ is fixed,
the various small excitations of the field $\Phi$ around $\Phi_-$
can be classified as massless or massive. The only components of
$\Phi$ that can be non-trivial in the kink solution are the
massive modes since the massless modes, also called the
Nambu-Goldstone modes, if non-vanishing inside the kink, will not
decay as we go further away from the kink. The massive modes are
given precisely by the generators that commute with $\Phi_-$ while
the Nambu-Goldstone modes are those that do not commute. Hence
$[\Phi_-, \Phi_k (x)]=0$ and, in particular, $[\Phi_-, \Phi_+]=0$.

Therefore to construct a kink solution, one needs to fix $\Phi_-$
to a vacuum expectation value and consider all possible commuting
vacuum expectation values for $\Phi_+$. $\Phi_-$ can be chosen to
be diagonal and by performing rotations that leave $\Phi_-$
invariant ({\it i.e.} lie in the unbroken group $H$ at
$x=-\infty$) $\Phi_+$ can also be brought to diagonal form.

Now we can explicitly list all the possible boundary conditions
(up to gauge rotations) that can lead to kink solutions. At
$x=-\infty$, we fix $\Phi_-=\Phi_0$ given in eq. (\ref{phi0}).
Then we can have
\begin{eqnarray}
\Phi_+&{}&=
 \epsilon_T \eta \sqrt{2 \over {N(N^2-1)}} \times \nonumber \\
       &{}& {\rm diag} ( n{\bf 1}_{n+1-q}, -(n+1){\bf 1}_{q},
                         n{\bf 1}_{q}, -(n+1){\bf 1}_{n-q}) \ ,
\label{phi+choices}
\end{eqnarray}
where we have introduced a parameter $\epsilon_T =\pm 1$ and
another $q=0,...,n$. The label $\epsilon_T$ is $+1$ when the
boundary conditions are topologically trivial and is $-1$ when
they are topologically non-trivial. $q$ tells us how many diagonal
entries of $\Phi_-$ have been permuted in $\Phi_+$. The case $q=0$
is when $\Phi_+ = \epsilon_T \Phi_-$. The case $q=n$ was
considered in detail in Ref. \cite{Vac01}.

\section{Kink solutions}
\label{kinksolutions}

We now find kink solutions for any allowed boundary conditions
$\Phi_\pm$. As a starting point we take the following ansatz:
\begin{equation}
\Phi_k = F_+ (x) {\bf M_+} + F_- (x) {\bf M_-} + g(x) {\bf M}\ ,
\label{kinkexplicit2}
\end{equation}
where
\begin{equation}
{\bf M}_+ =  {{\Phi_+ + \Phi_-}\over {2}} \ , \ \ {\bf M}_- =
{{\Phi_+ - \Phi_-}\over {2}} \label{M+M-} \ ,
\end{equation}
$g(\pm \infty)=0$ and $M$ is yet to be found.
Explicitly, for $\epsilon_T=-1$, we have
\begin{eqnarray}
{\bf M}_+ = \eta &N& \sqrt{1 \over {2N(N^2-1)}} \nonumber \\
&{}& {\rm diag} ( 0_{n+1-q}, {\bf 1}_q, -{\bf 1}_q, 0_{n-q} )
\label{M+} \ ,
\end{eqnarray}
\begin{eqnarray}
{\bf M}_- = &\eta&  \sqrt{1 \over {2N(N^2-1)}} \nonumber \\
&{}&{\rm diag} ( -2n {\bf 1}_{n+1-q}, {\bf 1}_q,
                {\bf 1}_q, 2(n+1){\bf 1}_{n-q} ) \ .
\label{M-}
\end{eqnarray}
Note that the matrices ${\bf M}_\pm$ are orthogonal:
\begin{equation}
{\rm Tr}({\bf M}_+{\bf M}_-) = 0 \ , \label{trM+M-}
\end{equation}
but are not normalized to 1/2. The boundary conditions for $F_\pm$
are:
\begin{eqnarray}
F_- (- \infty ) &=& -1 \ , \ \  F_- (+\infty ) =+1 \ , \nonumber \\
F_+ (- \infty ) &=& +1 \ , \ \  F_+ (+\infty ) =+1 \ .
\label{Fpmbc}
\end{eqnarray}
The advantage of this form of the ansatz is that, for particular
values of the parameters of a quartic potential in the $q=n$
topological ($\epsilon_T=-1$) case, one finds the explicit and
simple solution $F_- (x) = \tanh (\sigma x)$, $F_+ (x) =1$ and
$g(x)=0$, where $\sigma$ is the kink width which can be written in
terms of the parameters \cite{PogVac00,Vac01}. Also, for $q=0$,
$\epsilon_T=-1$, the solution is the embedded $Z_2$ kink {\it
i.e.} $F_+(x)=g(x)=0$, $F_- (x) = \tanh (\sigma x)$.

Now we would like to find the unknown matrix ${\bf M}$ in the ansatz
(\ref{kinkexplicit2}).  This can be done by treating $g(x) {\bf M}$
as a small perturbation to
\begin{equation}
\Phi_k^{(0)}\equiv F_+ (x) {\bf M_+} + F_- (x) {\bf M_-} \, .
\label{phi-try}
\end{equation}
The perturbation is restricted to generators that are
orthogonal to $\Phi_k^{(0)}$:
\begin{equation}
{\rm Tr}(\Phi_k^{(0)} {\bf M}) = 0  \ . \label{phikM}
\end{equation}
We need to check if the energy density contains any terms that are
linear in $g(x)$, otherwise we could always construct a stable
kink solution with $g(x)=0$. The quadratic terms in the energy
density clearly will not have such terms since ${\rm
Tr}(\Phi_k^{(0)} {\bf M}) =0$. The only terms that may be linear
in $g(x)$ will be from terms in the potential such as ${\rm Tr}
(\Phi^s )$ for even $s \ge 4$. ($s$ has to be even since the
potential is taken to have a $Z_2$ symmetry under $\Phi
\rightarrow -\Phi$.) There will be no terms linear in $g(x)$ only
if
\begin{equation}
{\rm Tr} ((\Phi_k^{(0)})^{s-1} {\bf M}) = 0 \label{trphikM}
\end{equation}
for every possible choice of ${\bf M}$ satisfying the conditions:
\begin{equation}
{\rm Tr} ({\bf M}) = 0 \ , \ \ {\rm Tr} ({\bf M}_- {\bf M}) = 0 \
, \ \ {\rm Tr} ({\bf M}_+ {\bf M}) = 0  \ . \label{conditionsonM}
\end{equation}
If ${\bf M}$ is off-diagonal, eq. (\ref{trphikM}) is satisfied
because the trace of the product of a diagonal and an off-diagonal
matrix vanishes. ($\Phi_k^{(0)}$ is diagonal.) The non-trivial
part is to check the condition for diagonal ${\bf M}$ and we shall
now concentrate on this case.

Let us write ${\bf M}$ as:
\begin{equation}
{\bf M} = {\rm diag} ({\bf U}_{n+1-q}, {\bf V}_q ,
                        {\bf W}_q, {\bf X}_{n-1}) \ ,
\label{Mmatrices}
\end{equation}
where ${\bf U}_{n+1-q}$, ${\bf V}_q$, ${\bf W}_q$ and ${\bf
X}_{n-1}$ are diagonal matrices of order given by their
subscripts. Implementation of the conditions in eq.
(\ref{conditionsonM}) leads to:
\begin{equation}
{\rm Tr} {\bf U}_{n+1-q} = - {\rm Tr} {\bf V}_q = -{\rm Tr} {\bf
W}_q = {\rm Tr} {\bf X}_{n-q}  \ . \label{conditionsonUVWX}
\end{equation}
Note that if $q=0$ or if $q=n$, this condition enforces each
matrix to be traceless.

Now, to check if eq. (\ref{trphikM}) is satisfied, we insert the
form of $\Phi_k^{(0)}$ from eq. (\ref{phi-try}). From
the boundary conditions in eq. (\ref{Fpmbc}), it is clear that the
functions $F_\pm (x)$ are linearly independent and so eq.
(\ref{trphikM}) can only be satisfied if:
\begin{equation}
{\rm Tr} ({\bf M}_+^\alpha {\bf M}_-^\beta {\bf M}) = 0
\label{conditiononta}
\end{equation}
for integers $\alpha$, $\beta$ such that $0 \le \alpha+\beta \le
s-1$. Explicit evaluation of this trace, together with the
relations in eq. (\ref{conditionsonUVWX}) shows that the condition
is satisfied by all ${\bf M}$ with ${\rm Tr}{\bf V}_q=0$. However,
for ${\bf M}$ with ${\rm Tr}{\bf V}_q\ne 0$, the condition is not
met if $\alpha$ is an even integer.

How many generators are there for which ${\rm Tr}{\bf V}_q \ne 0$
and that satisfy the conditions in eq. (\ref{conditionsonUVWX})?
There are a total number of $N-1$ diagonal $SU(N)$ generators. Of
these, the number of generators satisfying the conditions in eq.
(\ref{conditionsonUVWX}) together with ${\rm Tr}{\bf V}_q=0$ are
$$
(n+1-q-1)+(q-1)+(q-1)+(n-q-1)=N-4 .
$$
Hence there are $(N-1)-(N-4)=3$ choices of ${\bf M}$ for which the
condition in eq. (\ref{conditionsonUVWX}) plus ${\rm Tr}{\bf
V}_q=0$ is not met. However this number includes the two
possibilities ${\bf M}={\bf M}_\pm$. Hence there is only one
remaining possible choice of ${\bf M}$ and this is:
\begin{eqnarray}
{\bf M} = &\mu& \, {\rm diag} ( q(n-q){\bf 1}_{n+1-q}, \nonumber \\
   &-&(n-q)(n+1-q){\bf 1}_q , -(n-q)(n+1-q){\bf 1}_q, \nonumber \\
   && \hskip 1.0 truecm q(n+1-q){\bf 1}_{n-q} )
\label{Mresult}
\end{eqnarray}
with $\mu$ being a normalization factor in which we also include
the energy scale $\eta$ for convenience:
\begin{equation}
\mu = \eta [ 2q(n-q)(n+1-q)\{ 2n(n+1-q)-q\} ]^{-1/2} \ .
\label{muvalue}
\end{equation}
Note that the matrix ${\bf M}$ is not normalizable if $q=0$ or if
$q=n$. For these values of $q$, we can set $g(x)=0$ and
$\Phi_k^{(0)}$ coincides with the ansatz $\Phi_k$.

It is easy to see that $\Phi_k$ is a valid ansatz. Any
perturbations that are orthogonal to $\Phi_k$ would have to
satisfy eq. (\ref{conditionsonUVWX}) as well as be orthogonal to
${\bf M}$. Such perturbations necessarily have ${\rm Tr}{\bf V}_q
=0$. Further, all traces of the kind in eq. (\ref{trphikM}) are
proportional to ${\rm Tr}{\bf V}_q$ and hence vanish. This
justifies the ansatz in eq. (\ref{kinkexplicit2}).

The functions $F_\pm (x)$ and $g(x)$ can be found by solving their
equations of motion derived from the Lagrangian together with the
specified boundary conditions. There is no guarantee that a solution
will exist and so we find the solutions explicitly for $N=5$ with
a quartic potential in Sec. \ref{su5kinks}.

An interesting point to note is that the ansatz is valid even if
$\Phi_\pm$ are not in distinct topological sectors {\it i.e.} even
if $\epsilon_T=+1$. These imply the existence of non-topological
kink solutions in the model. If we include a subscript $NT$ to
denote ``non-topological'' and $T$ to denote ``topological'', we
have
\begin{equation}
\Phi_{NTk} = F_+ (x) {\bf M}_{NT+} + F_- (x) {\bf M}_{NT-}
          + g(x) {\bf M}_{NT} \ .
\label{NTfinalansatz}
\end{equation}
Since $\Phi_{NT+}=-\Phi_{T+}$, we find
\begin{equation}
{\bf M}_{NT+} = {\bf M}_{T-} \ , \ \ {\bf M}_{NT-} = {\bf M}_{T+}
\ , \ \ {\bf M}_{NT} ={\bf M}_{T} \ . \label{NTtoT}
\end{equation}
Hence
\begin{equation}
\Phi_{NTk} = F_- (x) {\bf M}_{T+} + F_+ (x) {\bf M}_{T-}
          + g(x) {\bf M}_T \ .
\label{NTsolution}
\end{equation}
So to get $F_-$ ($F_+$) for the non-topological kink we have to
solve the topological $F_+$ ($F_-$) equation of motion with the
boundary conditions for $F_-$ ($F_+$). To obtain $g$ for the
non-topological kink, we need to interchange $F_+$ and $F_-$ in
the topological equation of motion. The boundary conditions for
$g$ are unchanged.

In Sec. \ref{su5kinks} we will find the topological and the
non-topological kinks explicitly for $N=5$. Generally the
non-topological solutions, if they exist, will be unstable.
However, the possibility that some of them may be locally stable
for certain potentials cannot be excluded.

\section{Kink classes}
\label{kinkclasses}

In Sec. \ref{kinkbc} we showed that there is a discrete set of
boundary conditions that lead to different topological kink
solutions. The discrete set is labeled by the integer $q$ which
runs from $0$ to $n$. Hence there are $n+1$ distinct classes of
kink solutions in the $SU(N)\times Z_2$ model under consideration
\cite{Vac01}.

The explicit construction of the $n+1$ classes of kinks has
already been described in Sec. \ref{kinksolutions}. Eq.
(\ref{kinkexplicit2}) describes the form of the solution for a
fixed value of $q$. A solution of this form is one member of the
class of kinks labeled by $q$. What are the other members of the
class?

The members of a class of kinks is given by the set of boundary
conditions that will lead to gauge equivalent kinks. In other
words, there is a set of transformations belonging to the unbroken
symmetry group, $H_-$ in eq. (\ref{unbrokensymm}) defined by the
vacuum expectation value $\Phi_-$, that will leave $\Phi_-$
invariant but will rotate $\Phi_+$ non-trivially. The kink
solutions obtained by these global gauge transformation will appear
different from the original kink at the level of field
configurations but are degenerate and belong to the same class. If
$K_q$ is the subgroup of $H_-$ that leaves the $q$-kink
solution, $\Phi_k$, invariant, then
$$
\Sigma_q \equiv H_-/K_q
$$
describes the class of $q$-kinks.

Another way to describe $\Sigma_q$ is in terms of all perturbative
modes that do not change the energy of the solution {\it i.e.} the
zero modes on the solution background. This will include modes
that give spatial translations and internal space rotations. The
translations have not been included in $\Sigma_q$, while the
internal space rotations have been included just as in the case
of a ``moduli space''. However, the internal zero modes may not
vanish at $x=+\infty$ and hence are not required to be normalizable.

Now we will find $\Sigma_q$ for various $q$.

When $q=0$, $\Phi_k$ is proportional to $\Phi_-$ and $K_q =H_-$
{\it i.e.} the symmetry group that leaves the kink invariant is
the entire unbroken symmetry group. Therefore $\Sigma_0 = 1$ and
there is only one element in the $q=0$ kink class.

When $0< q <n$, it is clear from eq. (\ref{phi+choices}) that the
elements of $H_-$ that leave $\Phi_+$ invariant are $SU(n+1-q)$ in
the first block, $SU(q)$ in the second block, $SU(q)$ in the third
block, and $SU(n-q)$ in the fourth block. In addition, the
diagonal generators of $H_-$ commute with $\Phi_+$ and these yield
another three $U(1)$ factors. Hence the boundary condition at
$x=+\infty$ is invariant under
\begin{equation}
[SU(n+1-q) \times (SU(q))^2 \times SU(n-q) \times U(1)^3]/Z_K \ ,
\label{qkinksymm}
\end{equation}
where we have modded out the continuous group by its center,
symbolically denoted by $Z_K$. (This is necessary since the center
of $SU(n+1-q)$ for example, is also contained in the $U(1)$
factors.) From the form of ${\bf M}$ in eq. (\ref{Mresult}), it is
clear that the group in (\ref{qkinksymm}) is also the symmetry
group that leaves ${\bf M}$ invariant. Hence it is also the
symmetry group that leaves the entire kink solution $\Phi_k$
invariant and so:
\begin{eqnarray}
K_q = [SU(n+1-q) \times (SU(q))^2 &\times& SU(n-q)\nonumber \\
 &\times& U(1)^3]/Z_K \ .
\label{qkinksymmfinal}
\end{eqnarray}
Therefore $\Sigma_q = H/K_q$ where $H$ is given in eq.
(\ref{unbrokensymm}) and $K_q$ in eq. (\ref{qkinksymmfinal}).

When $q=n$, the analysis is modified a little bit since now
$n-q=0$ and the last block in $\Phi_+$ is absent. So now we have
\begin{equation}
K_n = [ (SU(n))^2 \times U(1)^2 ]/Z_K \ . \label{nkinksymm}
\end{equation}

Note that the above classification scheme holds for both
topological ($\epsilon_T=-1$) and non-topological
($\epsilon_T=+1$) kink solutions.

The space $\Sigma_q$ ($q\ne 0$) has interesting topological properties.
For example, it has a non-trivial second homotopy group. This suggests
that certain spherical configurations of domain walls (in three
spatial dimensions) will be topologically non-trivial and may not
be able to contract. We postpone a detailed investigation of the
interpretation of the non-trivial topology of $\Sigma_q$ and its
consequences for future work.

\section{Kink solutions for $N=5$}
\label{su5kinks}

In this section we will explicitly construct the kink solutions
when $N=5$ and when the potential is quartic:
\begin{equation}
V(\Phi ) = - m^2 {\rm Tr}[ \Phi ^2 ]+ h ( {\rm Tr}[\Phi ^2  ])^2 +
      \lambda {\rm Tr}[\Phi ^4]  + V_0 \ .
\label{quarticV}
\end{equation}
The desired symmetry breaking to
\begin{equation}
H= [SU(3)\times SU(2)\times U(1)]/[Z_3\times Z_2] \label{His321}
\end{equation}
is achieved in the parameter range
\begin{equation}
{h \over \lambda} > - {{N^2+3}\over {N(N^2-1)}}\biggl |_{N=5}
                     = - {7\over {30}} \ .
\label{symmbreakparam}
\end{equation}
The vacuum expectation value, $\Phi_-$ is
\begin{equation}
\Phi_- = \eta {1\over \sqrt{60}} (2,2,2,-3,-3) \label{su5phi-}
\end{equation}
with
\begin{equation}
\eta \equiv {{m} \over {\sqrt{\lambda '}}} \label{eta}
\end{equation}
and
\begin{equation}
\lambda ' \equiv
      h +  {{N^2+3}\over {N(N^2-1)}}\biggl |_{N=5} \lambda
      = h + {7\over {30}} \lambda \ .
\label{lambdaprime}
\end{equation}

The $q=0$ topological kink ($\Phi_+=-\Phi_-$) has been found in
Ref. \cite{PogVac00} and is simply an embedded $Z_2$ kink for
all parameters:
\begin{equation}
\Phi_k^{q=0} =  \tanh \left ( {{mx}\over {\sqrt{2}}} \right )
                  \Phi_- \ .
\label{su5qis0}
\end{equation}
As discussed in Sec. \ref{kinkclasses}, there is only one kink
solution in this class.

\begin{figure}
\vskip 0.5 truecm
\epsfxsize= 0.95\hsize\epsfbox{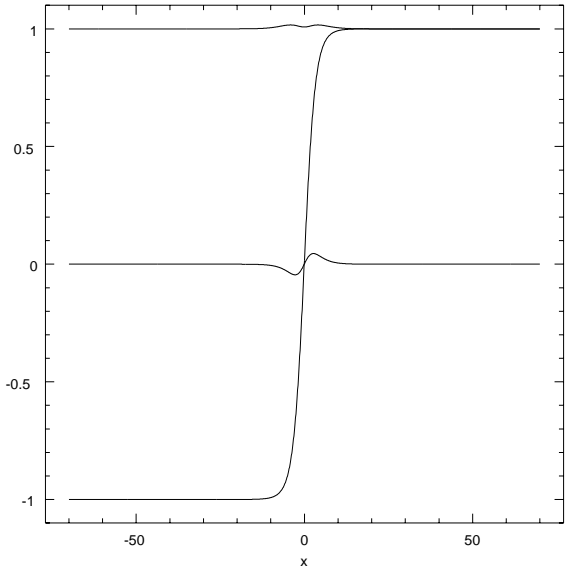}
\vskip 0.5 truecm
\caption{\label{qeq1top} The profile functions $F_+$ (nearly
1 throughout), $F_-$ (shaped like a $\tanh$ function), and $g$
(nearly zero) for the $q=1$ topological kink with parameters
$h=-3/70$, $\lambda=1$ and $\eta =1$.}
\end{figure}

To find the $q=1$ topological kink solution, we use the ansatz
found in Sec. \ref{kinksolutions}
\begin{equation}
\Phi_k^{q=1} = F_+{\bf M}_+ + F_- {\bf M}_- + g {\bf M}
\end{equation}
with
\begin{equation}
{\bf M}_+ = \eta \sqrt{5\over {48}} {\rm diag}(0,0,1,-1,0) \ ,
\label{bfM+su5}
\end{equation}
\begin{equation}
{\bf M}_- = \eta {1\over {\sqrt{240}}} {\rm diag}(-4,-4,1,1,6) \ ,
\label{bfM-su5}
\end{equation}
\begin{equation}
{\bf M} = \eta {1\over {2\sqrt{7}}} {\rm diag}(1,1,-2,-2,2) \ .
\label{bfMsu5}
\end{equation}
Inserting the ansatz in the Lagrangian we can derive the equations
of motion for the functions $F_\pm$ and $g$. (These are given in
Appendix \ref{appendixb}.) The boundary conditions on these
functions are:
\begin{equation}
F_+ (\pm \infty )=1\ , \ \ F_- (\pm \infty ) = \pm 1 \ , \ \ g(\pm
\infty ) = 0 \ . \label{su5bcs}
\end{equation}
If we assume that $|g''| << m^2 |g| << 1$ and $|F_+''| << m^2 |F_+|$,
an approximate analytic solution can be obtained when
$h = -3\lambda /70$. (The assumptions can later be
checked for self-consistency.) The approximate solution is:
\begin{equation}
F_- \simeq \tanh \left ( {m\over {\sqrt{2}}}  x \right ) \ ,
\label{approxF-}
\end{equation}
\begin{equation}
g \simeq - {{\gamma_6 F_- (\alpha_1 + \alpha_2 F_-^2)} \over
          {(\alpha_2 \gamma_1 -\alpha_1 \gamma_3) +
     (\alpha_2 \gamma_4 + \alpha_5 \gamma_6) F_-^2}} \ ,
\label{approxg}
\end{equation}
\begin{equation}
F_+ \simeq \alpha_2^{-1/2}
      [ -\alpha_1 - \alpha_5 g F_- ]^{1/2} \ ,
\label{approxF+}
\end{equation}
where the coefficients $\alpha_i$ and $\gamma_i$ are
given in Appendix \ref{appendixb}. This approximate solution
can be extended to other near-by parameters and a comparison
with the numerically obtained solutions shows that the
approximation is reasonably good except at the turning
points of $F_+$ and $g$. However, the qualitative features of
the numerical solution are captured by the approximation.
We show the numerical solution for $h=-3\lambda /70$ in
Fig. \ref{qeq1top}. A numerical investigation for other values
of $h/\lambda$ shows that a solution always exists for the $q=1$
topological kink.

The class of $q=1$ kinks is described by the space
\begin{equation}
\Sigma_1 = H/K_1 \label{sigma1} \ ,
\end{equation}
where
\begin{equation}
K_1 = [SU(2) \times U(1)^3]/Z_2 \label{k1forsu5} \ .
\end{equation}

The $q=2$ kink has been found in Ref. \cite{PogVac00} (also see
\cite{Vac01}). In the case when
\begin{equation}
{h\over \lambda} = - {3\over {20}} \label{hlambdaqeq2}
\end{equation}
the solution can be written down simply as:
\begin{equation}
\Phi_k^{q=2} = {{1-\tanh(\sigma x)}\over 2} \Phi_- +
                {{1+\tanh(\sigma x)}\over 2} \Phi_+
\label{qeq2solution}
\end{equation}
with
\begin{equation}
\Phi_+ = - \eta {1\over \sqrt{60}} (2,-3,-3,2,2)  \ .
\label{qeq2su5phi+}
\end{equation}
($\Phi_-$ is given by eq. (\ref{su5phi-}) and $\sigma
=m/\sqrt{2}$.)

A more general ansatz, valid for all values of $h/\lambda$, is
\begin{equation}
\Phi_k^{q=2}={{F_+(x)-F_-(x)}\over 2} \Phi_- +
                {{F_+(x)+F_-(x)}\over 2} \Phi_+ \, ,
\label{q2ansatz}
\end{equation}
where functions $F_+$ and $F_-$ satisfy the same boundary
conditions as in (\ref{su5bcs}). The equations of motion for the
$q=2$ kink along with a numerical solution were presented in
\cite{PogVac00}.

\vskip -0.25 truein

The class of $q=2$ kinks is described by the space
\begin{equation}
\Sigma_2 = H/K_2 \label{sigma2} \ ,
\end{equation}
where
\begin{equation}
K_2 = [SU(2)^2 \times U(1)^2]/Z_2^2 \ . \label{k2forsu5}
\end{equation}

Now we will also construct the non-topological ($\epsilon_T=+1$)
kinks in the model.

The $q=0$ non-topological kink is simply the
vacuum $\Phi_{NTk} = \Phi_+$ and there is only one member in this
class.

\begin{figure}
\vskip 0.5 truecm \epsfxsize= 0.95\hsize\epsfbox{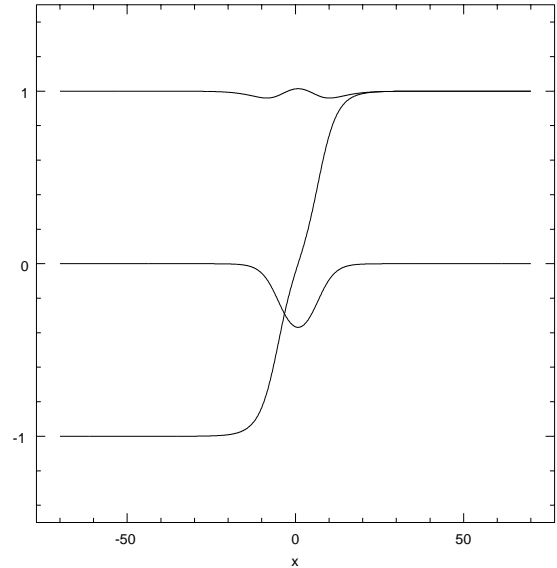} \vskip
0.5 truecm \caption{\label{q1nontop14o70} The profile functions
$F_+$ (shaped like a $\tanh$ function), $F_-$ (nearly constant at
$1$), and $g$ (asymptotically zero) for the $q=1$ non-topological
kink with parameters $h=-14/70$, $\lambda=1$ and $\eta =1$.}
\end{figure}

As discussed at the end of Sec.  \ref{kinksolutions}, to construct
the $q=1$ non-topological kink we can use the same equations as
for the topological case but we should switch the boundary
conditions on $F_+$ and $F_-$ (eq. (\ref{su5bcs})). The system of
equations has been solved numerically for a few choices of
parameters. For $h=-14\lambda /70$, the profile functions are
shown in Fig. \ref{q1nontop14o70}. For $h=-3\lambda /70$ we find
that the $q=1$ non-topological kink breaks up into two $q=2$
topological kinks. Specifically the $q=1$ kink interpolating
between $\Phi \propto (2,2,2,-3,-3)$ and $(2,2,-3,2,-3)$ breaks up
into one $q=2$ kink interpolating between $(2,2,2,-3,-3)$ and
$-(-3,-3,2,2,2)$ and another interpolating between
$-(-3,-3,2,2,2)$ and $(2,2,-3,2,-3)$ This suggests that there is a
repulsive force between different $q=2$ kinks for parameters close
to $h=-3\lambda /70$ and so there will be no non-topological $q=1$
kink solution in a certain range of parameters. Numerically we
have determined the critical parameter where the $q=1$
non-topological boundary conditions lead to two well-separated
topological $q=2$ kinks instead of one bound object. Hence we find
that there are no $q=1$ non-topological kink solutions for $h >
-0.18 \lambda$ .

\vskip -0.25 truein

The $q=2$ non-topological kink can be found by solving the same
equations of motion as for the topological $q=2$ kink after
switching the boundary conditions on $F_+$ and $F_-$ ($g=0$ in
this case). Then, for the parameter $h=-3\lambda/20$, one has
\begin{equation}
\Phi_{NTk}^{q=2} = {{1-\tanh(\sigma x)}\over 2} \Phi_- +
                {{1+\tanh(\sigma x)}\over 2} \Phi_+ \ ,
\label{NTqeq2solution}
\end{equation}
where
\begin{equation}
\Phi_+ = + \eta {1\over \sqrt{60}} (2,-3,-3,2,2)  \ .
\label{NTqeq2su5phi+}
\end{equation}
For general values of parameters the profile functions can be
found by numerical relaxation.


\section{Kink stability}
\label{kinkstability}

To analyze the stability of the various kink solutions, we have to
expand the energy density to second order in perturbations and
then look for unstable modes. This would have to be done on a case
by case basis for every different choice of potential. Here we
will analyze the stability of the $SU(5)$ kinks constructed in the
previous section.

The $q=0$ topological kink is known to be unstable
\cite{PogVac00}. To see this, note that the Nambu-Goldstone modes
are massless at $x=\pm \infty$ and have a negative mass squared at
the origin where $\Phi_k =0$. Furthermore,
it can be checked that the mass squared for the
Nambu-Goldstone modes is everywhere negative for any choice of
parameters. We know that an
everywhere negative potential in one dimension always admits a
bound state. Therefore the $q=0$ topological kink is unstable
towards the growth of the Nambu-Goldstone modes for all parameters.

The $q=1$ topological kink is perturbatively unstable. The
unstable modes correspond to the four generators of $SU(5)$ which
commute with $\Phi^{q=1}_k (0) \propto {\bf M}_+$ and do not commute
with $\Phi_-$ and $\Phi_+$.
These modes are massless at $x=\pm \infty$ and have a non-zero
mass at the origin. The corresponding potential is given by
\begin{eqnarray}
\nonumber U^{q=1}(x) = &-& m^2+{7\over 12}(h+{2\lambda\over
5})\eta^2 F_-^2 + {5\over 12} \eta^2 h F_+^2 \\ &+&
(h+{\lambda\over 2})\eta^2 g^2 + \sqrt{7 \over 60} \eta^2 \lambda
F_- g \, . \label{q1potential}
\end{eqnarray}
We have evaluated $U^{q=1}(x)$ numerically and found that it is
everywhere negative for any choice of parameters.

As shown in Ref. \cite{PogVac00}, the $q=2$ topological kink is
perturbatively stable, at least for a range of parameters around
the choice in eq. (\ref{hlambdaqeq2}).

Next we discuss the perturbative stability of non-topological
kinks.

The $q=0$ non-topological kink is simply the vacuum and is
trivially stable.

\begin{figure}
\vskip 0.5 truecm \epsfxsize= 0.95\hsize\epsfbox{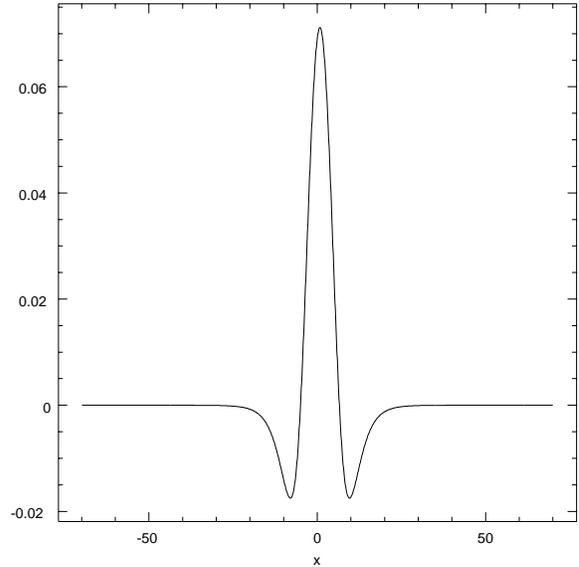}
\vskip 0.5 truecm \caption{\label{fig:potq1nt} $U_{NT}^{q=1}(x)$
versus $x$ for $h=-14\lambda/70$ with $\lambda=1$ and $\eta
=1$.}
\end{figure}

We have seen that the $q=1$ non-topological kink solution may not
exist for some parameter values. In other words, the $q=1$
configuration may split and become two $q=2$ topological kinks.
When the $q=1$ non-topological kink does not split into two
well-separated $q=2$ topological kinks, we find that it is locally
stable. The potentially unstable modes are the two generators of
$SU(5)$ that commute with $\Phi_k^{q=1NT}(0) \propto {\bf M}_-$
and do not commute with $\Phi_-$ and $\Phi_+$. The corresponding
potential has a particularly simple form:
\begin{equation}
U_{NT}^{q=1}(x) = {F_+'' \over F_+} \, . \label{q1ntpotential}
\end{equation}
The plot of $U_{NT}^{q=1}(x)$ versus $x$ for $h/\lambda=-14/70$
is shown in Fig. \ref{fig:potq1nt}. We have checked that the value
of the potential at $x=0$ remains positive for all parameters for
which the $q=1NT$ kink solution exists.

\vskip -0.25 truein

The $q=2$ non-topological kink is perturbatively unstable for all
parameter choices. The unstable modes are the eight
Nambu-Goldstone modes for which the potential is given by the same
expression as in eq. (\ref{q1ntpotential}). Numerically we find
that $U_{NT}^{q=2}(x) < 0$ for all $x$.

A general statement we can make is that the topological kinks in
one of the classes will be globally stable. This just follows from
the fact that the kinks are topological and so there must be a
lowest energy kink. In the analysis done for the $SU(5)$ case in
Sec. \ref{su5kinks}, the $q=n$ kink is the least energetic while
the $q=0$ kink has the largest number of unstable modes. This
suggests that perhaps the $q=n$ topological kink is the globally
stable kink for any choice of potential and not just the quartic
potential considered in this section. Another argument in
support of this conjecture is that the change in the values of the
field components in going from $x=-\infty$ to $+\infty$ is the
least for the $q=n$ kink. Only one component need vanish inside
the core of the $q=n$ kink while a greater number of components
vanish inside the core for $q\le n-1$. The situation with the
non-topological kinks is precisely the opposite. Here we know that
the $q=0$ non-topological kink is the vacuum and hence is the
least energy state.

\section{$SU(5)$ magnetic monopoles}
\label{su5mm}

A possible ansatz for a spherically symmetric $SU(5)$
fundamental magnetic monopole solution is \cite{DokTom80,WilGol77}:
\begin{equation}
\Phi_M \equiv \sum_{a=1}^{3} P(r) {\hat r}^a T^a + M(r) T^4 +
                                     N(r) T^5 \ ,
\label{monopolesolution}
\end{equation}
where the subscript $M$ denotes the monopole field configuration,
\begin{eqnarray}
T^a = \frac{1}{2}{\rm diag}(0,0,\sigma^a,0)\ &,& \ \
T^4 = \frac{1}{2\sqrt{3}}(1,1,0,0,-2) \ , \nonumber \\
T^5 = {1\over {2\sqrt{15}}}&(&2,2,-3,-3,2) \ ,
\label{Tagenerators}
\end{eqnarray}
$\sigma^a$ are the Pauli spin matrices, $r=\sqrt{x^2+y^2+z^2}$
is the spherical radial coordinate, and ${\hat r}^a$ denotes the
unit radial vector. The ansatz for the gauge fields for the monopole
can also be written down
$$
W^a_i = \epsilon^a_{~ij}
\frac{{\hat r}^j}{er}(1-K(r))  \ , \
(a=1,2,3) \ ,
$$
\begin{equation}
W^b_i = 0 \ , \ \ (b \ne 1,2,3) \ , \label{gaugeansatz}
\end{equation}
where $e$ is the gauge coupling. $P(r)$, $M(r)$, $N(r)$ and $K(r)$
are profile functions.

In the BPS case, when the $SU(5)$ potential vanishes,
the exact, minimal energy solution is known \cite{Mec99}:
\begin{equation}
P(r) = \frac{1}{er}(\frac{Cr}{\tanh(Cr)}-1) \ , \
K(r) = \frac{Cr}{\sinh(Cr)} \ ,
\label{bpsphiprofile}
\end{equation}
\begin{equation}
M(r) = \frac{2}{\sqrt{3}}\frac{C}{e} \ , \ \
N(r) = {1\over {\sqrt{15}}}\frac{C}{e} \ .
\label{bconmandn}
\end{equation}
where $C$ is a constant.

We can also write the monopole aysmptotic field configuration in more
transparent form as
$\Phi_M (r=\infty ) = U^\dagger_{34} \Phi_+ U_{34}$ where
$$
U_{34} (\theta , \phi )=
        e^{-i\phi T^3} e^{-i\theta T^2} e^{+i\phi T^3} \ ,
$$
$\theta$, $\phi$ are spherical angular coordinates and the
generators $T^a$ are given in eq. (\ref{Tagenerators}).
Note that the winding of the monopole lies entirely in
the $(3,4)$ block of $\Phi$.
We are now -- in contrast to the earlier sections -- also choosing
\begin{equation}
\Phi_+ = \eta {1\over \sqrt{60}} (2,2,2,-3,-3) \ .
\end{equation}
Any other choice can be transformed to this choice by a global
$SU(5)$ rotation.

The existence of the BPS solution does not preclude the existence of
other higher energy magnetic monopole solutions even for fixed asymptotics
since the boundary conditions at the origin can be chosen in different
ways. (Ansatze with other asymptotics can be found in
\cite{DokTom80}.) One possible route to determining
the different monopole boundary conditions at $r=0$ is to assume
that the cores of magnetic monopoles are like the cores
of domain walls. Then we would like to find the different spherical
domain walls that have the asymptotics of the BPS solution. This will
provide the spherical domain walls with monopole topology.
If these spherical domain
walls can shrink to zero size, the collapse will produce a monopole
whose core is the same as that of the spherical domain wall that we
started out with. In this way we might hope to determine the
different possibilities for the boundary condition $\Phi_M (0)$.

We have three classes $q=0,1,2$ each of topological and non-topological
walls. Let us consider each of these classes one by one.

The $q=0NT$ ($q=0$, non-topological) kink is trivial and we need not
discuss it any further. The $q=0T$ ($q=0$, topological) kink has
$$
\Phi^{q=0}_k (x) = \tanh (\sigma x) \Phi_+ \ .
$$
Using the kink solution, we can write down a field configuration
corresponding to a spherical $q=0T$ domain wall:
$$
\Phi^{q=0T} (r, \theta ,\phi ) \approx \tanh (\sigma (r-R)) \Phi_+
\ ,
$$
where $R$, the radius of the spherical domain wall, is taken to be
very large. Next we would like to introduce monopole topology as a
boundary condition to get an object that is a monopole in which
all the energy resides in a shell made of a domain wall. (We call
this object a ``monopole-wall'' (MW).) To do this
we need to apply an $SU(5)$ rotation $U_{34}$ on $\Phi$. This
will generally be ill-defined at the center ($r=0$) of the spherical
domain wall since the field there will then become multi-valued.
However, we are ultimately interested in letting the radius of
the spherical domain wall go to zero and hence we need only
apply the gauge transformation on $\Phi$ for $r\ge R$.
Therefore $\Phi$ for the monopole-wall is:
$$
\Phi_{MW}^{q=0T} (r,\theta ,\phi) \approx \tanh (\sigma (r-R))
              U^\dagger_{34} \Phi_+ U_{34} \ , \ \ \ r > R \ .
$$
Note that the value of the field in the core of the wall
is the same everywhere on the wall, that is,
$\Phi_{MW}^{q=0} (R,\theta , \phi)=0$
regardless of the spherical angular coordinates.
Therefore the monopole-wall can collapse to a point and the
field will remain single-valued. The resulting monopole
will have $\Phi_M (r=0) = 0$. That is, the new boundary conditions
on $M(r)$ and $N(r)$ suggested by this argument are: $M(0)=0=N(0)$.

Next consider the $q=1NT$ kink.
Here $\Phi^{q=1NT} (0) \propto (4,4,-1,-1,-6)$ in the core of
the domain wall. Once again we may construct the monopole-wall by
applying the transformation $U_{34}$. Since
$$
U^\dagger_{34} \Phi^{q=1NT} (0) U_{34} \propto \Phi (0) \ ,
$$
the monopole-wall can collapse into a monopole. This suggests
that we should be able to find a monopole solution with
$\Phi_M (r=0) \propto (4,4,-1,-1,-6)$. This is precisely the
monopole with boundary conditions given in eq. (\ref{bconmandn}).

The $q=1T$ kink has
$\Phi^{q=1T} (0) \propto (0,0,1,-1,0)$ and this is not invariant
under rotations by $U_{34}$. Therefore once we impose monopole
boundary conditions on a spherical domain wall of this type, the
field in the core of the domain wall will depend on the angular
coordinates. Such a wall cannot simply collapse to zero radius
since that would violate single-valuedness of the field. Hence
we do not expect to find a monopole whose center has $\Phi$
proportional to $(0,0,1,-1,0)$.

The $q=2T$ kink has
$\Phi^{q=2T} (0) \propto (0,1,1,-1,-1)$ and, as this is not
invariant under $U_{34}$, a monopole with
$\Phi_M (0) \propto (0,1,1,-1,-1)$ is not possible.

The $q=2NT$ kink as described in Sec. \ref{su5kinks} has
$\Phi^{q=2NT} (0) \propto (-4,1,1,1,1)$ and this is invariant
under $U_{34}$. This suggests that a monopole with $\Phi_M (0)
\propto (-4,1,1,1,1)$ is possible. However, this monopole-wall
does not quite fit the form of the monopole solution given in eq.
(\ref{monopolesolution}). We find that if we choose $M(0) =
-\sqrt{5} N(0)$, the center of the monopole has $\Phi_M (0)
\propto (1,1,1,1,-4)$ and not $(-4,1,1,1,1)$. A global $SU(5)$
rotation on the monopole solution could be used to make $\Phi_M
(0) \propto (-4,1,1,1,1)$, however this would then rotate the
asymptotic field to $\Phi_M (z=\infty ) \propto (-3,2,2,-3,2)$,
once again providing a mismatch between the monopole-wall and the
monopole ansatz in eq. (\ref{monopolesolution}). In spite of this
mismatch, the monopole-wall has the same topologically non-trivial
asymptotic field configuration as the BPS solution and can also
contract to a point without any conflict with single-valuedness.
Hence we think that a monopole solution with $\Phi_M (0) \propto
(-4,1,1,1,1)$ should exist.

The above discussion, suggesting that there could be several
monopole solutions corresponding to different boundary
conditions on the scalar field at $r=0$, clearly applies to 
global monopoles. In the case of gauge monopoles, the only 
non-trivial gauge fields are the three fields associated with 
the $SU(2)$ group of the embedded monopole, as in the BPS case 
above. These fields still satisfy the form in eq. (\ref{gaugeansatz}) 
and the only quantity that will depend on the ``monopole generation'' 
is the profile function $K(r)$.

This completes an analysis of all the cases. Three of the five
non-trivial cases led to the possibility of a monopole solution.
This suggests the existence of three classes of fundamental
monopoles in $SU(5)$ with the same asymptotics as the
BPS monopole.

\section{Conclusions}
\label{conclusions}

We have shown that the kink solutions in $SU(N)\times Z_2$ occur
in $(N+1)/2$ classes. All the kink solutions, regardless of class,
have the same topological charge. Borrowing the terminology of the
standard model where particles come in ``generations'' (or
``families''), we dub the kink classes ``kink generations''. We
have determined the continuous degeneracy associated with every
kink generation. The degeneracy is described by certain manifolds
which themselves have interesting topological properties. In
particular, the manifolds have non-trivial second homotopy,
suggesting that certain configurations of closed domain walls in
three spatial dimensions may be incontractable.

We have also examined the stability of the various classes of
kinks in an $SU(5)$ model with quartic potential. Our analysis
shows that two classes of solutions are perturbatively stable
(for some parameters) while the other non-trivial kinks are
unstable.

The generation structure of domain walls suggests a generation
structure for the magnetic monopoles in the gauged version of the
model - a possibility that seems worth exploring further in the
context of the dual standard model \cite{Vac96}. We have found
that spherical domain walls of the $q=0T, 1NT, 2NT$ classes can
collapse into monopoles that al have the same asymptotic field
configurations. Hence monopole solutions with $\Phi_M (0)=0$ and
$\Phi_M(0) \propto (-4,1,1,1,1)$ should be possible to construct
in addition to the known case where $\Phi_M (0) \propto
(4,4,-1,-1,-6)$. If all these different boundary conditions lead
to magnetic monopole solutions and there are none
others\footnote{While the only $SU(5)$ monopole solution known to
us is the BPS solution, an exhaustive list of spherically
symmetric ansatze consistent with monopole topology is given in
\cite{DokTom80}. Some of these could possibly lead to other
monopole solutions with different asymptotic field
configurations.}, it would indicate that there are exactly three
generations of $SU(5)$ magnetic monopole solutions. To confirm
this statement would require an explicit construction of the
$SU(5)$ monopole solutions with the various possible boundary
conditions.

We anticipate that a survey of the space of $SU(N)$ magnetic monopole
solutions will show novel features, similar to those we have discovered
in the case of kinks.

\acknowledgements

Conversations with Gautam Mandal and Spenta Wadia are gratefully
acknowledged. This work was supported by the DoE.

\appendix

\section{Proof that solutions require $[\Phi_+,\Phi_-]=0$}
\label{appendixa}

Let $\Phi_k (x)$ be a kink solution. We can expand the solution in
an orthonormal set of $SU(N)$ generators $T^a$ (${\rm Tr} (T_a
T_b) = \delta_{ab}/2$):
\begin{equation}
\Phi_k (x) = \sum_a \phi_a (x) T^a \ . \label{Phi0expansion}
\end{equation}
Here an alternate expansion will be more convenient:
\begin{equation}
\Phi_k (x) = \sum_a \psi_a (x) R^a \ ,
\label{Phi0alternateexpansion}
\end{equation}
where
\begin{equation}
R^1 \equiv {1\over \eta} \Phi_- \equiv R_- \ , \ \ \ R^2 \equiv
{1\over \eta} \Phi_+ \equiv R_+  \ , \label{R1R2definition}
\end{equation}
where $\eta$ is a normalization factor so that ${\rm Tr} (R_\pm
^2) = 1/2$ and the remaining $R^a$ complete the set of generators.
Depending on the boundary conditions, it may well turn out that
${\rm Tr}(R_+ R_- ) \ne 0$ and so these generators are not
orthogonal. However, we shall choose the other generators, {\it
i.e.} $R^a$ with $a \ne 1,2$, to satisfy the orthogonality
conditions ${\rm Tr} (R_+ R^a) = 0 = {\rm Tr} (R_- R^a)$ and also
normalize them to satisfy ${\rm Tr} (R_a R^a)=1/2$. We define new
structure constants $r_{abc}$ by
\begin{equation}
[ R^a , R^b ] = i r_{abc} R^c \ . \label{rstructureconstants}
\end{equation}

Next we need to state certain properties of the functions $\psi_a
(x)$. Due to the boundary conditions $\Phi (x\rightarrow \pm
\infty ) \rightarrow \Phi_\pm$, we have
\begin{equation}
\psi_1 (-\infty ) = \eta \ , \ \ \psi_a (-\infty ) = 0 \ (a \ne 1)
\ ,  \label{minusinfinity}
\end{equation}
\begin{equation}
\psi_2 (+\infty ) = \eta \ , \ \ \psi_a (+\infty ) = 0 \ (a \ne 2)
\ . \label{plusinfinity}
\end{equation}
(Just as for the generators, $\psi_- \equiv \psi_1$ and $\psi_+
\equiv \psi_2$.) These boundary conditions ensure that there is no
non-trivial solution of the kind $\psi_a (x) = {\rm constant}$.

Let us now perturb the kink solution $\Phi_k (x)$. For this,
consider the field configuration
\begin{equation}
\Phi_1 (x) = U(x) \Phi_k U^\dagger (x) \label{Phi1} \ ,
\end{equation}
where $U(x) \in SU(N)$. Note that $V(\Phi_1 )=V(\Phi_k)$ since the
potential is invariant under $SU(N)$ local gauge transformations.
Then the energy of the configuration $\Phi_1$ is:
\begin{eqnarray}
E[\Phi_1] = E[\Phi_k ]  +
 2 {\rm Tr}( \partial_x \Phi_k [&&U^\dagger \partial_x U, \Phi_k ]
 )                            \nonumber \\
 && + {\rm Tr} ( [U^\dagger \partial_x U ,\Phi_k ]^2 ) \ .
\label{energy}
\end{eqnarray}
If we now consider infinitesimal rotations, the second term is
linear in these while the last term is quadratic. If $\Phi_k$ is
to be a solution, the linear variation must vanish. Therefore,
\begin{equation}
{\rm Tr}( \partial_x \Phi_k [U^\dagger \partial_x U, \Phi_k ] ) =0
\label{solutioncondition}
\end{equation}
for all $U(x)$ infinitesimally close to unity and for all $x$.

The condition in eq. (\ref{solutioncondition}) can also be
rewritten as:
\begin{equation}
{\rm Tr}( [\Phi_k , \partial_x \Phi_k ] U^\dagger \partial_x U )
=0 \ , \label{conditionrewrite}
\end{equation}
which should hold for any $U(x) \in SU(N)$. (For infinitesimal
rotations this condition is
\begin{equation}
{\rm Tr}( [\Phi_k , \partial_x \Phi_k ] T^a ) =0 \ , \ \forall x,
a \ , \label{smallconditionrewrite}
\end{equation}
where $T^a$ form a complete set of $SU(N)$ generators.) Hence the
solution must necessarily satisfy
\begin{equation}
[\Phi_k , \partial_x \Phi_k ] = 0 \label{commcondition}
\end{equation}
for all $x$.

Next use the expansion of $\Phi_k$ of eq.
(\ref{Phi0alternateexpansion}) in eq. (\ref{commcondition}) and
that gives us:
\begin{equation}
\sum_{b>a} r_{abc}
    [\psi_a (x) \psi_b '(x) - \psi_b (x) \psi_a ' (x) ] = 0 \ ,
\ \forall c, x  \ .\label{condition2}
\end{equation}

If the functions
$$
F_{ab} \equiv \psi_a (x) \psi_b '(x) - \psi_b (x) \psi_a ' (x)
$$
are linearly independent, eq. (\ref{condition2}) implies that
$r_{abc} =0$ whenever $F_{ab}\ne 0$. It is easy to see that
$F_{ab}\ne 0$ provided both $\psi_a$ and $\psi_b$ are non-trivial
and linearly independent. Hence the (assumed) linear independence of
$F_{ab}$ implies that $r_{abc}=0$ whenever $\psi_a$ and $\psi_b$ are
non-trivial and linearly independent. It is sufficient to assume
that all the $\psi_a$ are linearly independent since if two
components are linearly dependent, the basis of generators, $R^a$,
can be redefined so that only linearly independent functions occur
in the expansion in eq. (\ref{Phi0alternateexpansion}). This shows
that if $F_{ab}$ are linearly independent then $[R^a, R^b] =0$ if
$\psi_a$ and $\psi_b$ are non-trivial. Therefore the solution
$\Phi_k$ can be expanded in a Cartan basis and in particular
$[\Phi_+, \Phi_-]=0$.

Without assuming the linear independence of the functions
$F_{ab}$, we can still show the desired result $[\Phi_\pm ,
\Phi_k]=0$ by examining the condition in eq. (\ref{condition2}) as
$x\rightarrow +\infty$. In this spatial region, the only
non-vanishing function is $\psi_+ (x) \rightarrow \eta$. The term
$\psi_+ \psi_a '$ is small because all derivatives vanish at
infinity. The terms $\psi_a \psi_b'$ with $\psi_a \ne \psi_+$ are
also small since both $\psi_a$ and $\psi_b '$ tend to zero at
$x=+\infty$. In this region, where the field is nearly at its
vacuum value, we can examine the behavior of the fields by
perturbing the potential around the vacuum. This tells us that
$\psi_a$ ($a\ne +$) falls off exponentially as $x \rightarrow
\infty$. Therefore,
$$
\psi_+ \psi_a ' >> \psi_a \psi_b'
$$
for all $a\ne +$. So the condition in eq. (\ref{condition2}) in
the large, positive $x$ region yields
\begin{equation}
\sum_{b\ne 2} r_{2bc} \psi_b '(x)  = 0 \ . \label{condition3}
\end{equation}
An integration over the interval $(x,+\infty )$ then gives
\begin{equation}
\sum_{b\ne 2} r_{2bc} \psi_b (x)  = 0  \ , \label{condition4}
\end{equation}
where we have used the boundary conditions $\psi_b (+\infty )=0$
except for $b=2$ (which does not appear in the sum). As discussed
above, it is sufficient to consider the case when the set of
functions $\psi_b (x)$ are linearly independent. Therefore, if
$\psi_b$ is non-trivial, we get
\begin{equation}
r_{2bc} = 0 \ , \ \ \ \forall b , c \ . \label{r2b1}
\end{equation}
Similarly, by considering the region with $x \rightarrow -\infty$,
\begin{equation}
r_{1bc} = 0 \ , \ \ \ \forall b , c \ . \label{r1b2}
\end{equation}
This shows that $[R_+ , R^a]=0=[R_- , R^a]$ if $\psi_a \ne 0$ for
any choice of $a$ and hence $[\Phi_\pm , \Phi_k (x)]=0$. In
particular, we can only get a kink solution if $[R_+,R_-]=0$ which
is equivalent to $[\Phi_+,\Phi_-]=0$.

\section{Equations of motion for the $\lowercase{q}=1$ kink in $SU(5)$}
\label{appendixb}

The equations of motion for the topological $q=1$ kink functions
$F_\pm$ and $g$ are:
\begin{eqnarray}
-F_+''+ \alpha_1 F_+ + \alpha_2 F_+^3 &+& \alpha_3 F_+ F_-^2
  +\alpha_4 g^2 F_+
\nonumber \\
    &+&\alpha_5 gF_- F_+=0 \ ,
\label{F+eq}
\end{eqnarray}
\begin{eqnarray}
-F_-''+ \beta_1 F_- &+& \beta_2 F_-^3 + \beta_3 F_+^2 F_-
  + \beta_4 g^2 F_-
\nonumber \\
        &+&\beta_5 g(3F_-^2-F_+^2) +\beta_6 g^3  =0 \ ,
\label{F-eq}
\end{eqnarray}
\begin{eqnarray}
-g''+\gamma_1 g &+&\gamma_2 g^3 +\gamma_3 gF_+^2
              + \gamma_4 gF_-^2
                                         \nonumber \\
   &+& \gamma_5 g^2F_- + \gamma_6 F_-(F_-^2-F_+^2) =0 \ ,
\label{geq}
\end{eqnarray}
where
\begin{eqnarray}
\alpha_1 &=& \beta_1 = \gamma_1 = -m^2 \, ,
\nonumber \\
\alpha_2 &=& \eta^2 {5\over 12} (h + {1 \over 2}\lambda) \, ,
\nonumber \\
\alpha_3 &=& \eta^2 {7\over 12} (h + {3 \over 70}\lambda) \, ,
\nonumber \\
\alpha_4 &=& \eta^2 (h + {6 \over 7}\lambda) \, ,
\nonumber \\
\alpha_5 &=& - \eta^2 \lambda \sqrt{3 \over 35} \, ,
\nonumber \\
\beta_2 &=& \eta^2 {7\over 12} (h + {181 \over 490}\lambda) \, ,
\nonumber \\
\beta_3 &=& \eta^2 {5\over 12} (h + {3 \over 70}\lambda) \, ,
\nonumber \\
\beta_4 &=& \eta^2 (h + {138 \over 245}\lambda) \, ,
\nonumber \\
\beta_5 &=& \eta^2 \lambda {5 \over 14} \sqrt{3 \over 35}\, ,
\nonumber \\
\beta_6 &=& \eta^2 \lambda {12 \over 49} \sqrt{3 \over 35}\, ,
\nonumber \\
\gamma_2 &=& \eta^2 (h + {25 \over 98}\lambda) \, ,
\nonumber \\
\gamma_3 &=& \eta^2 {5 \over 12} (h + {6 \over 7}\lambda) \, ,
\nonumber \\
\gamma_4 &=& \eta^2 {7 \over 12}(h + {138 \over 245}\lambda) \, ,
\nonumber \\
\gamma_5 &=& \eta^2 \lambda {3 \over 7} \sqrt{3 \over 35}\, ,
\nonumber \\
\gamma_6 &=& \eta^2 \lambda {5 \over 24} \sqrt{3 \over 35}\, .
\end{eqnarray}


\begin{thebibliography}{}

\bibitem{PogVac00}
L. Pogosian and T. Vachaspati, Phys. Rev. {\bf D62}, 123506
(2000).

\bibitem{Vac01}
T. Vachaspati, Phys. Rev. {\bf D63}, 105010 (2001).

\bibitem{DokTom80}
C. P. Dokos and T. N. Tomaras
Phys. Rev. {\bf D21}, 2940 (1980)

\bibitem{WilGol77}
D. Wilkinson and A. Goldhaber,
Phys. Rev. {\bf D16}, 1221 (1977).

\bibitem{Mec99}
M. Meckes, private communication (1999).

\bibitem{Vac96}
T. Vachaspati, Phys. Rev. Lett. {\bf 76}, 188 (1996).

\end{thebibliography}
\end{document}